\documentclass[doublecol,linenumbers]{epl2}
\usepackage{amsmath}
\usepackage{graphics}

\title{Molecular Beam Epitaxy Growth of Superconducting LiFeAs Film on SrTiO$_3$(001) Substrate}

\author{K. Chang\inst{1,2} \and P. Deng\inst{1,2} \and T. Zhang\inst{1,2} \and H.-C. Lin\inst{1,2} \and K. Zhao\inst{1,2} \and S.-H. Ji\inst{1,2}\thanks{E-mail: \email{shji@mail.tsinghua.edu.cn}} \and L.-L. Wang\inst{1,2} \and K. He\inst{1,2} \and X.-C. Ma\inst{1,2} \and X. Chen\inst{1,2}\thanks{E-mail: \email{xc@mail.tsinghua.edu.cn}} \and Q.-K. Xue\inst{1,2}\thanks{E-mail: \email{qkxue@mail.tsinghua.edu.cn}}}
\shortauthor{K. Chang \etal}

\institute{
  \inst{1} State Key Laboratory of Low-Dimensional Quantum Physics, Department of Physics, Tsinghua University, Beijing 100084, China\\
  \inst{2} Collaborative Innovation Center of Quantum Matter, Beijing 100084, China
}
\pacs{81.15.Hi}{Molecular beam epitaxy}
\pacs{74.70.Xa}{Iron pnictide  superconductor}
\pacs{68.37.Ef}{Scanning tunneling microscopy}

\abstract{
The stoichiometric ``111'' iron-based superconductor, LiFeAs, has attacted great research interest in recent years. For the first time, we have successfully grown LiFeAs thin film by molecular beam epitaxy (MBE) on SrTiO$_3$(001) substrate, and studied the interfacial growth behavior
by reflection high energy electron diffraction (RHEED) and low-temperature scanning tunneling microscope (LT-STM).
The effects of substrate temperature and Li/Fe flux ratio were investigated.
Uniform LiFeAs film as thin as 3 quintuple-layer (QL) is formed. Superconducting gap appears in LiFeAs films thicker than 4~QL at 4.7~K.
When the film is thicker than 13~QL, the superconducting gap determined by the distance between coherence peaks is about 7~meV,
close to the value of bulk material. The \textit{ex situ} transport measurement of thick LiFeAs film shows a sharp superconducting transition
around 16~K. The upper critical field, $H_{c2}(0)=13.0$~T, is estimated from the temperature dependent magnetoresistance.
The precise thickness and quality control of LiFeAs film paves the road of growing similar ultrathin iron arsenide films.}

\begin{document}

\maketitle

\section{Introduction}

\begin{figure*}
\onefigure[width=6.6in]{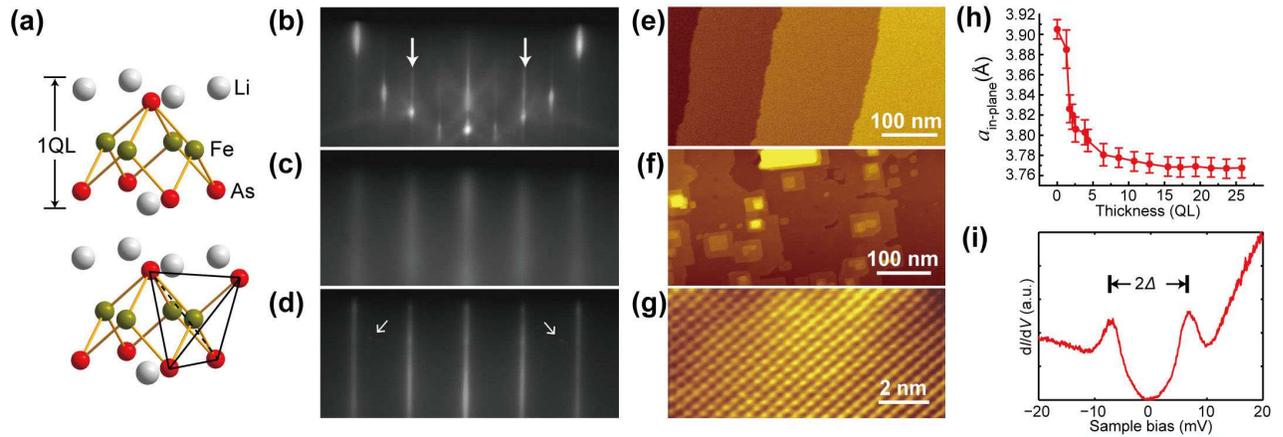}
\caption{(Color online) MBE growth of LiFeAs film.
(a) The tetragonal structure of LiFeAs. (b) RHEED pattern of (001) surface of SrTiO$_3$ substrate, which shows the 2$\times$2 reconstruction.
The arrows indicate the 1$\times$1 stripes. (c,d) RHEED patterns of LiFeAs thin film on SrTiO$_3$ substrate after 8 (c) and 60 (d) minutes deposition.
The weak spots marked by the arrows show the formation of small amount of Li$_3$As clusters on the surface during growth.
(e,f) STM topographic images of SrTiO$_3$(001) (e) and 25 QL LiFeAs film (f). Image size: 500~nm$\times$250~nm, sample bias $1.0$~V,
tunneling current $100$~pA. (g) Atomically resolved STM topography image on the surface of a 25~QL LiFeAs film
(10~nm$\times$10~nm, $8.0$~mV, $2.0$~nA). (h) The evolution of the in-plane lattice constant $a_{\text{in-plane}}$ determined by RHEED
as a function of the film thickness. (i) The $dI/dV$ spectrum on the surface of a 13~QL LiFeAs film.
}
\label{fig1}
\end{figure*}

LiFeAs is a ``111'' type iron-based superconductor (Fe-SC)
and has attracted much attention recently due to its unusual properties compared to other superconducting iron pnictides.
Without chemical doping, the stoichiometric LiFeAs already shows a superconducting transition at 18~K under the ambient pressure \cite{jin08,chu08,pitcher08}.
More interestingly, the magnetic structure driven by the Fermi surface nesting between electron-like and hole-like pockets,
an essential ingredient for the parent compound of iron pnictide superconductors, is absent in LiFeAs \cite{buechner10,takahashi12,felner09,buechner12}.
Both $s_{\pm}$-wave singlet and $p$-wave triplet pairing have been proposed for LiFeAs \cite{platt11,brydon11}.

LiFeAs has a layered tetragonal structure $(p4/nmm)$.
As shown in fig.~\ref{fig1}(a), an edge-sharing FeAs$_4$-tetrahedra layer in LiFeAs consists of one layer of Fe sandwiched between two layers of As.
Compared with the square lattice formed by the Fe atoms, the lattice of As atoms rotates by 45$^\circ$ and enlarges by a factor of $\sqrt{2}$ \cite{jin08,chu08,pitcher08}.
Two layers of Li atoms are intercalated between FeAs and form the neutral cleaving plane without surface reconstruction.
Therefore, the surface sensitive probes, such as scanning tunneling microscope (STM) and angle resolved photoelectron spectroscopy (ARPES),
have been successfully applied to capture the intrinsic properties of LiFeAs \cite{davis12,pennec12,hess12,hanaguri12,buechner10,takahashi12}.

Synthesis of high quality single-crystal is critical for a comprehensive study of LiFeAs.
Similar to some other iron pnictides, LiFeAs was first synthesized by the high pressure method \cite{jin08,chu08,pitcher08}.
The flux and self-flux methods have also been introduced to produce larger size single crystal up to millimeter \cite{kwon10,buechner10b}. There has been no LiFeAs film grown by MBE, a powerful method to grow single crystal thin films up to centimeter scale, till now.
Although several other iron arsenide superconductors have been grown by MBE \cite{agatsuma10,ueda11,ueda12,takeda12}, those films were all thicker than 100~nm and the interfacial behavior has never been studied. In order to achieve  thickness control with atomic precision, we performed epitaxial growth of LiFeAs single crystal thin films on SrTiO$_3$(001) surface.
The superconductivity in LiFeAs film is then characterized by STM and transport measurement.

\begin{figure*}
\onefigure[width=6.6in]{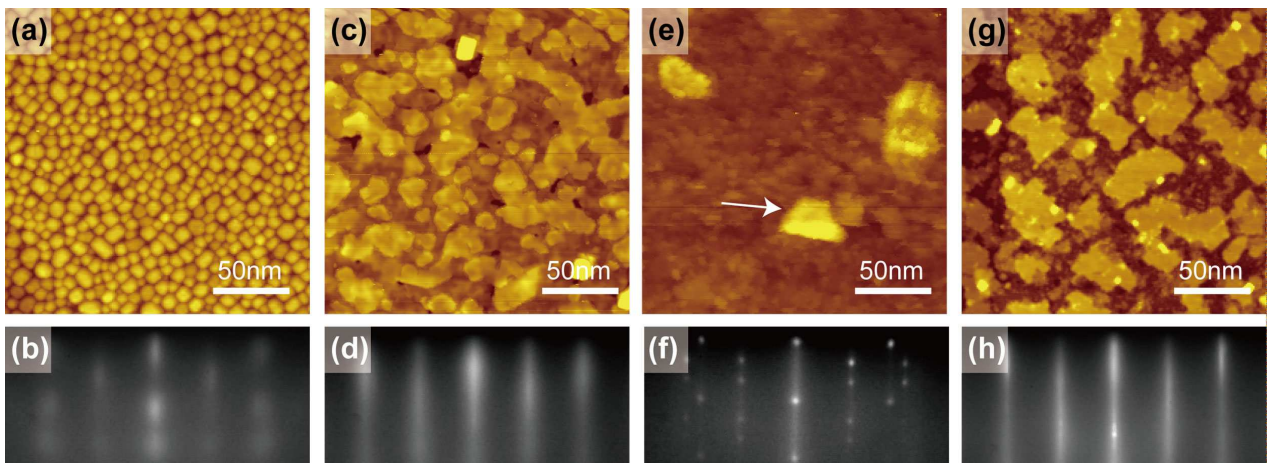}
\caption{(Color online)
Evolution of topography and RHEED pattern with different Li flux and substrate temperature.
(a-f) STM images and RHEED of LiFeAs films grown at 345~$^\circ$C substrate temperature with the same Fe flux.
The Li source temperatures are 374~$^\circ$C (a,b), 390~$^\circ$C (c,d) and 400~$^\circ$C (e.f), respectively.
The arrow in (e) indicates a triangular Li$_3$As island. (g,h) STM image and RHEED of a sample
grown at 450~$^\circ$C substrate temperature. All the fluxes are the same as (e).
Image conditions for all STM images: $3.0$~V and $30$~pA.
        }
\label{fig2}
\end{figure*}

\begin{figure}
\onefigure[width=3.25in]{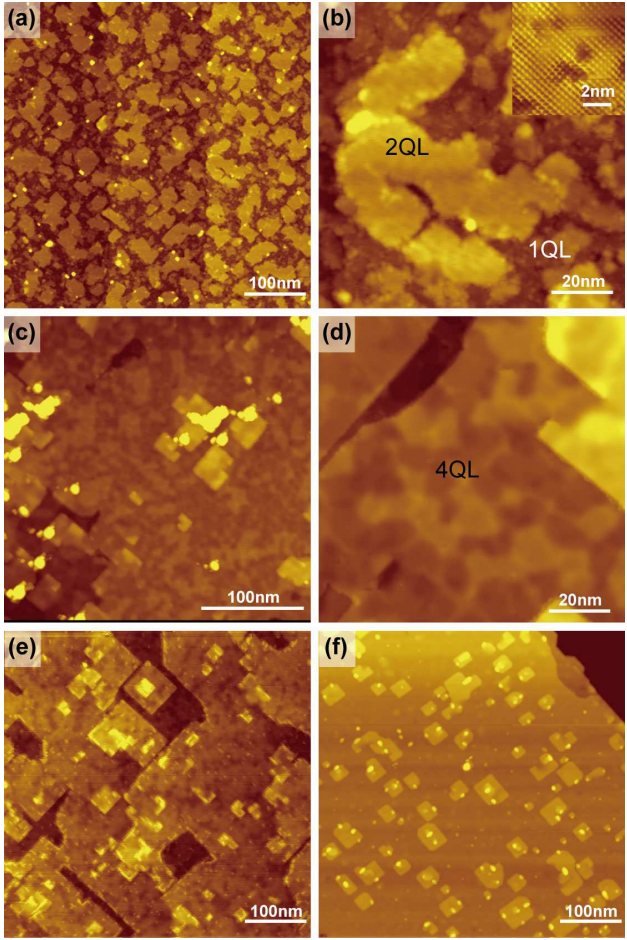}
\caption{(Color online)
STM topography of LiFeAs films with various thicknesses.
The temperature of substrate was fixed at 450~$^\circ$C during growth.
 (a) 1.5~QL (500~nm $\times$~500 nm, $3.0$~V, $30$~pA).
(b) Detailed features of the film in (a) (100~nm $\times$~100 nm, $2.0$~V, $50$~pA). Inset, the atom resolved image obtained from a 2QL island (8~nm $\times$~8 nm, $50$~mV, $100$~pA).
(c) 4.5~QL (300~nm $\times$~300 nm, $3.0$~V, $30$~pA).
(d) Detailed features of the film in (c) (100~nm $\times$~100 nm, $1.0$~V, $80$~pA).
(e) 7.5~QL (500~nm $\times$~500 nm, $3.5$~V, $25$~pA). (f) 13~QL (500~nm $\times$ 500~nm, $4.0$~V, $30$~pA).
        }
\label{fig3}
\end{figure}

\begin{figure}
\onefigure[width=3.25in]{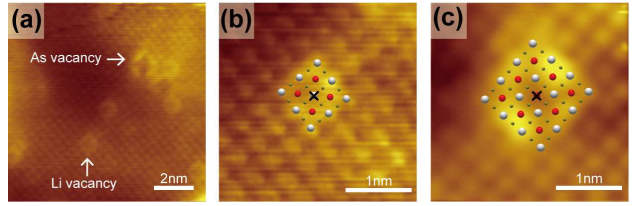}
\caption{(Color online)
Defects in LiFeAs film. (a) Image showing Li and As vacancies. $5.0$~mV, $1.0$~nA.
(b) A Li vacancy. $5.0$~mV, $1.0$~nA.
(c) An As vacancy. $8.0$~mV, $0.4$~nA.
The lattice model of LiFeAs is superposed on the images.
White, red and grey balls stand for the topmost Li, As and Fe atoms, respectively.
The vacancy is indicated by a cross.
        }
        \label{fig4}
\end{figure}

\begin{figure}
\onefigure[width=3.25in]{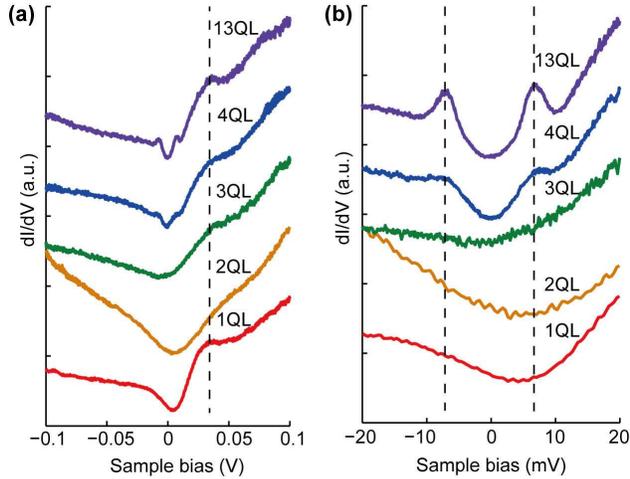}
\caption{(Color online)
Thickness dependence of the $dI/dV$ spectra. (a) from $-0.1$~V to 0.1~V and (b) from $-20$~mV to 20~mV. The spectra are vertically shifted for clarity.
Superconducting gap begins to emerge when thickness is 4~QL. The dashed line in (a) indicates the kink at 36~meV,
which was also seen in the cleaved LiFeAs single crystals.
The dashed lines in (b) label the coherence peaks.
        }
        \label{fig5}
\end{figure}

\section{Experimental method}

The experiment was carried out on a Unisoku low-temperature STM combined with MBE.
The base pressure is lower than $1\times10^{-10}$ Torr. The LiFeAs films were grown in the MBE chamber
and immediately transferred into the STM at a temperature of 4.7~K without taking out of the ultra-high vacuum (UHV) environment.
A PtIr alloy tip was used in STM for topography and spectroscopy measurement.

The Nb-doped SrTiO$_3$(001) substrate was cleaned in UHV by direct current heating up to 1200~$^\circ$C for several times.
Each time the heating lasted for 5 minutes.
The temperature of substrate was monitored by an infrared pyrometer. Li and Fe were evaporated from two standard Knudsen cells (K-cell).
The evaporation temperatures were around 400~$^\circ$C and 1100~$^\circ$C, respectively.
In the conventional MBE growth of arsenide, As flux is obtained by directly heating solid As, producing As$_2$ and As$_4$ molecules with low reactivity.
In order to obtain the more reactive atomic As, we used FeAs compound with high purity as the source for As.
FeAs compound was heated to $\sim400~^\circ$C in K-cell and decomposed into atomic Fe and As.
Due to the huge vapor pressure difference between As ($10^{-4}$ Torr at 200~$^\circ$C) and Fe ($10^{-4}$ Torr at 1220~$^\circ$C),
the flux consists of nearly pure As atoms, which was confirmed by quadruple mass spectrometer.
During the epitaxial growth, the flux of As was much higher than that of Fe (flux ratio $>$ 20:1).
The flux ratio between Fe and Li was finely tuned to prevent the formation of Li$_3$As or FeAs, which shows additional spots in the RHEED pattern.
A Sigma Q-pod crystal oscillator measured the flux of Fe.
The growth rate of LiFeAs films is approximately 0.2$\sim$0.4 quintuple-layer (QL, one LiFeAs unit cell) per minute, depending on the temperature of Fe source.

The transport measurements were carried out {\it ex situ} on a Quantum Design PPMS system, where the lowest temperature is 1.8~K and the highest magnetic field is 9~T.
To avoid the degrading of LiFeAs film in the ambient atmosphere, several nanometer thick Ag protection layer was deposited on the top of the LiFeAs epitaxial film before taking out of UHV.

\section{Results and discussion}

The sharp patterns and Kikuchi lines of reflection high energy electron diffraction (RHEED) in fig.~\ref{fig1}(b) show the cleanness and flatness
of the SrTiO$_{3}$ substrate
after high temperature treatment in UHV. Besides the 1$\times$1 pattern for the (001) surface along the $\langle010\rangle$ direction
[indicated by the arrows in fig.~\ref{fig1}(b)], the lines between the 1$\times$1 streaks show the 2$\times$2 reconstruction of the surface.

450~$^\circ$C is the optimal substrate temperature to achieve layer-by-layer growth of LiFeAs on SrTiO$_{3}$.
Upon growth, the RHEED pattern of substrate disappeared quickly and then much more broadened 1$\times$1 pattern [fig.~\ref{fig1}(c)]
appeared with the same orientation as the substrate, indicating the coherent epitaxial nature of the growth.
As the deposition time increases, the RHEED pattern becomes shaper. No RHEED intensity oscillation has been observed.
Figure 1(d) shows the 1$\times$1 RHEED pattern of a 25~QL LiFeAs.
The single-crystalline nature of the thin film was further verified by the RHEED patterns with different substrate azimuth.

The very weak spots marked by the arrows in fig.~\ref{fig1}(d) are attributed to a small amount of Li$_{3}$As clusters on the surface of the film,
which is introduced by slightly excessive Li flux. The superconductivity of LiFeAs films is not significantly influenced by these Li$_{3}$As clusters
(however excessive Fe flux does severely suppress superconductivity). With careful tuning of the Li flux, the extra spots can be completely removed.

In the STM image of the substrate [fig.~\ref{fig1}(e)], the average terrace width is up to several hundred nanometers.
The step height of 3.9~{\AA} corresponds to one unit cell of SrTiO$_{3}$.
The image in fig.~\ref{fig1}(f) reveals the atomically flat surface of an epitaxial 25~QL LiFeAs film on SrTiO$_3$ and the flat area size can be up to 100~nm.
The atomic step height 6.3~{\AA}, measured from the STM image, corresponds to one QL of LiFeAs.
In addition, rectangular islands and screw dislocations are visible on the terraces.

Atomically resolved image [fig.~\ref{fig1}(g)] shows the square lattice with an atomic spacing of 3.8~{\AA},
which is consistent with the lattice constant of LiFeAs(001) \cite{davis12, pennec12, hess12, hanaguri12}.
Similar to the cleaved (001) surface, no surface reconstruction has been observed on the epitaxial LiFeAs thin film.
We believe that the surface is terminated by Li instead of As since the As-terminated surface normally has reconstruction for iron pnictide superconductor \cite{davis10}.
The thickness dependence of the lattice constant is estimated from RHEED pattern and shown in fig.~\ref{fig1}(h).
The in-plane lattice constant of LiFeAs thin film continuously decreases from 3.91~{\AA}, the lattice constant of SrTiO$_{3}$ substrate, to 3.77~{\AA}, that of the LiFeAs bulk material.
The curve in fig.~\ref{fig1}(h) shows that the lattice of the first few epitaxial layers expands due to the lattice mismatch between LiFeAs and SrTiO$_3$ substrate.
The lattice of the film relaxes to that of the bulk material within 15~QL.
Well-defined superconducting gap with sharp coherence peaks is revealed by scanning tunneling spectroscopy (STS) on a film thicker than 13~QL [fig.~\ref{fig1}(i)].
The gap size $\Delta\sim7$~meV is consistent with the previous reports on bulk materials \cite{davis12,pennec12,hanaguri12}.

Li/Fe flux ratio is essential for LiFeAs film growth. Figures 2(a) to 2(f) illustrate three samples grown under
the same Fe flux and substrate temperature (345~$^\circ$C), but with different Li flux.
The nominal thickness of the samples is 3~QL. When Li is insufficient,
small islands of several nanometers appear [fig.~\ref{fig2}(a)] and the RHEED pattern [fig.~\ref{fig2}(b)] exhibits dim spots.
If the Li flux is increased to a certain value, continuous film [fig.~\ref{fig2}(c)] with high density of screw dislocations is formed.
The RHEED pattern [fig.~\ref{fig2}(d)] shows stripes corresponding to the LiFeAs lattice constant 3.8~{\AA}.
Further increase of Li flux leads to triangular islands on the film [fig.~\ref{fig2}(e)].
In the RHEED pattern [fig.~\ref{fig2}(f)], a set of sharp spots superposes on the LiFeAs stripe.
Analysis shows that the spots correspond to these triangular Li$_3$As islands on the surface.

Besides the Li/Fe flux ratio, substrate temperature is another important factor to determine the film topography.
In particular, the effective Li flux changes with the substrate temperature because of the desorption of Li atoms,
which is more pronounced when the Li source temperature (usually 350$\sim$400~$^\circ$C) is close to that of the substrate.
For example, the sample in fig.~\ref{fig2}(g) was grown under the same flux of Li, Fe and As as that in fig.~\ref{fig2}(e), but with
the substrate temperature raised to 450~$^\circ$C. There is no longer Li$_3$As islands on the film
and the RHEED shows similar pattern as that in fig.~\ref{fig2}(d) without the sharp spots.
Therefore, the Li source temperature has to be adjusted
for each substrate temperature to maintain effectively the same Li/Fe flux ratio.

By adjusting the Li/Fe flux ratio at the optimal substrate temperature (450~$^\circ$C),
we can obtain atomically flat film as thin as $\sim$3~QL.
Below 3~QL, LiFeAs islands with several tens of nanometer lateral size are formed [figs. 3(a) and 3(b)]. The atom resolved image on a 2QL island shows the square lattice constant of 3.85~\AA, in consistent with the RHEED data [fig. 3(b), inset].
The $dI/dV$ spectra measured off the islands are identical to that of the SrTiO$_3$ substrate,
confirming the island thickness assignment.
On SrTiO$_3$, it is difficult to follow the layer-by-layer growth mode, mainly because Li tends to form clusters on the substrate.
Thicker film becomes flatter and more uniform [fig.~\ref{fig3}(c)]. Screw dislocations appear on the surface, probably originating from
the merging of islands.
We note that the density of screw dislocation in the film grown at 450~$^\circ$C is much lower than that grown at 350~$^\circ$C.
The surface corrugation of a 4~QL film [fig.~\ref{fig3}(d)] is about 2~{\AA},
which may be attributed to the unreleased strain induced by the substrate.
When the film thickness increases to 7.5~QL, single QL islands appear  on the surface [fig.~\ref{fig3}(e)].
Above 13~QL [fig.~\ref{fig3}(f)], there are very few screw dislocations and very low surface corrugation,
implying that the strain has been sufficiently released.
The stain relief is also indicated by the lattice constant measured from RHEED patterns [fig.~\ref{fig1}(h)].

Two types of defects are visible on the surface [fig.~\ref{fig4}(a)]. The Li vacancy [fig.~\ref{fig4}(b)] appears as a hole in the topmost layer.
The eight neighboring Li atoms are brighter in the STM image. The larger defect [fig.~\ref{fig4}(c)] can be ascribed to the As vacancy in the second layer.
An As atom sits in the center of four Li atoms, and both the nearest and the next-to-nearest Li atoms in the topmost layer
are influenced by the absence of an As atom.
Interestingly, the Fe vacancies with 2-fold symmetry, which are frequently seen in the cleaved bulk materials \cite{pennec12b},
are seldom seen in the MBE film. In MBE growth, relatively low substrate temperature prevents the generation of Fe vacancies.
For comparison, the synthesis temperature for LiFeAs single crystal is usually as high as 800$^\circ$C \cite{jin08}.

We performed STS measurements on the films with different thicknesses [fig.~\ref{fig5}].
A superconducting gap at the Fermi level begins to emerge when the thickness is 4~QL. The 13~QL film shows a gap with pronounced coherence peaks.
The peak-to-peak distance is $2\Delta\sim14$~meV. The $dI/dV$ spectra of the film is consistent with that of the bulk materials \cite{hanaguri12},
including the kink at 36~meV [fig.~\ref{fig5}(a), indicated by dashed line].
Superconductivity is absent in a film thinner than 4~QL at liquid helium temperature, even after long time annealing.
Our result is similar with the FeSe film grown on the SiC(0001) substrate covered by epitaxial graphene \cite{songcl11}, in which superconductivity only exists in the films thicker than 2 triple-layers (TL), and $T_c$ changes gradually from 3.7~K to the bulk value with the increase of thickness. This is a normal behavior in superconductors and can be interpreted by adding a surface-energy term in the Ginzburg-Landau free energy \cite{simonin86}. The lateral finite size effect is excluded for the islands thicker than 2~QL, because the island size is much larger than the in-plane coherence length \cite{kurita11}. The tensile strain in the first few QLs of our LiFeAs film is less likely to be responsible to the absence of superconductivity because multiple high-pressure experiments proved that decreasing the lattice constant would suppress superconductivity \cite{gooch09,zhang09,mito09}.

It is interesting to compare our LiFeAs film with the FeSe film epitaxially grown on SrTiO$_3$(001) substrate. Although as grown 1~TL FeSe film is not superconducting, it shows superconducting $T_c$ as high as $\sim$50~K after proper annealing procedure \cite{wangqy12}. However, the second layer of FeSe is always insulating. ARPES experiments ascribed the superconductivity in 1~TL FeSe to the higher level of charge transfer from the substrate \cite{hesl13,liux14}. In our LiFeAs film, Li atomic layer is believed to be the interfacial layer with SrTiO$_3$ substrate due to the charge neutral consideration. Different from 1~TL FeSe, the electrons have to cross the Li layer to diffuse into FeAs layer, which decreases the transfer efficiency. Besides, part of the intercalated Li atoms are evaporated during the annealing process, further lowering the doping level. The island size of 1~QL LiFeAs, which is comparable with the coherence length, is another possible reason to the absence of superconductivity.

%\revision{It is interesting to compare our LiFeAs film with the FeSe film epitaxially grown on SrTiO$_3$(001) substrate. Although the as grown 1~TL FeSe film is not superconducting, it shows superconducting $T_c$ as high as $\sim$50~K after proper annealing procedure \cite{wangqy12}. However, the second layer of FeSe is always insulating. ARPES experiments ascribed this difference to the amount of charge transfer. With sufficient electron transfer from the substrate, the hole pocket at the $\overline\Gamma$ point vanishes and superconductivity emerges in 1~TL FeSe \cite{hesl13}. In contrast, the charge transfer is insufficient in the second layer of FeSe and the hole pocket at $\overline\Gamma$ always exists \cite{liux14}. In our LiFeAs film, Li atomic layer is believed to be the interfacial layer with SrTiO$_3$ substrate due to the charge neutral consideration. Different from 1~TL FeSe, the electrons have to cross the Li layer to diffuse into FeAs layer, which decreases the transfer efficiency. Besides, part of the intercalated Li atoms are evaporated during the annealing process, further lowering the doping level. Another possible reason is the relatively small size of the 1QL islands, which is comparable with the coherence length.}

\begin{figure*}
\onefigure[width=6.6in]{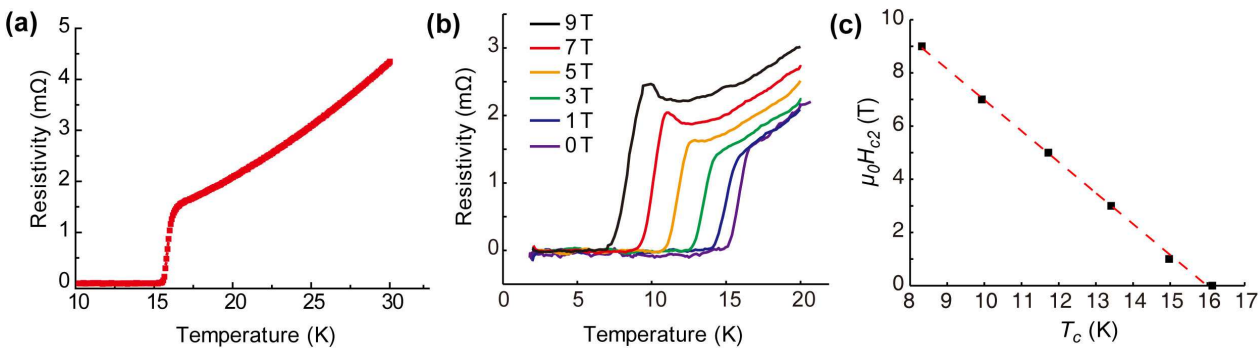}
\caption{(Color online)
(a) Temperature dependence of resistivity of $\sim$100~QL LiFeAs film with silver capping layer under zero magnetic field. It shows superconducting transition temperature $T_c=16$~K. (b) Magnetoresistance measured under the field perpendicular to the film from 0~T to 9~T. (c) Magnetic field dependence of the superconducting transition temperature $T_c$. The linear fitting gives the slope $dH_{c2}/dT=-1.17$~T/K, corresponding to critical field $H_{c2}=13.0$~T at zero temperature.
        }
        \label{fig6}
\end{figure*}

Furthermore, we measured the transport properties of the LiFeAs thin film.
Since LiFeAs degrades in the ambient atmosphere, several nanometer thick silver was deposited on top of LiFeAs
 as the protection layer before the sample was taken out of UHV for transport measurement.
 Figure 6(a) shows the temperature dependence of resistance for a 100~QL LiFeAs film with silver capping layer.
 The sharp superconducting transition occurs at 16~K, which is close to the bulk value of 18~K \cite{jin08,chu08,pitcher08}.
 Figure 6(b) shows the magnetoresistance with magnetic field perpendicular to the sample.
 The superconducting transition shifts to lower temperature as the magnetic field increases from 0~T to 9~T.
 At the same time the normal state resistance increases. The magnetic field dependence of $T_{c}$ [fig.~\ref{fig6}(c)]
 is almost linear between 0~T and 9~T. A linear fitting gives the slope $\mu_0dH_{c2}/dT=-1.17$~T/K. Calculating from the Werthamer-Helfand-Hohenberg formula $\mu_0H_{c2}(0)=-0.693\frac{\mu_0dH_{c2}(T)}{dT}|_{T_c}T_c$ \cite{WHH66},
  the upper critical field at zero temperature is 13.0~T, slightly lower than the bulk value 17 T \cite{cho11}.
 For LiFeAs film thinner than $\sim$50~QL, it is difficult to detect the superconducting transition
 since the film is unstable in atmosphere even with the silver capping layer.

\section{Conclusion}

We have realized the molecular beam epitaxial growth of LiFeAs thin film with high quality on SrTiO$_{3}$(001) substrate.
The MBE film makes it possible to study the intrinsic properties of LiFeAs.
At 4.7~K, the superconducting gap is revealed in ultra-thin film thicker than 4~QL.
The gap size of 13~QL film is as large as 7~meV, which is close to the bulk value.
The 100~QL film shows sharp superconducting transition at 16~K in transport measurement.
The upper critical field $H_{c2}$ is estimated to be 13.0~T from the magnetoresistance measurement.
The current growth approach can be applied to other materials consisting of highly reactive alkaline elements,
such as NaFeAs and KFe$_2$As$_2$.
Our work paves the way for further studies on the intrinsic properties of LiFeAs,
which can be disturbed by the impurities introduced in the bulk growth methods.

\acknowledgments
The authors thank C. Q. Jin for synthesizing the FeAs compound.
The work was financially supported by Ministry of Science and Technology of China under grant number 2013CB934600, National Science Foundation and Ministry of Education of China.

\end{document}